\documentstyle[11pt,epsfig]{article}
\title{\bf  How can rainbow gravity affect on gravitational force?}
\author{A. S. Sefiedgar\thanks{e-mail: a.sefiedgar@umz.ac.ir} 
\\ {\small Department of Physics, Faculty of Basic Sciences, University of Mazandaran}\\ {P.O. Box 47416-95447, Babolsar, Iran}}
\begin{document}
\maketitle 
\begin{abstract}
According to Verlinde's recent proposal, the gravity is originally an entropic force. In this work, we obtain the corrections to the entropy-area law of black holes within rainbow gravity. The corrected entropy-area law leads to the modifications of the number of bits $N$. Inspired by Verlinde's argument on the entropic force, and using the modified number of bits, we can investigate the effects of rainbow gravity on the modified Newtonian dynamics, Newton's law of gravitation,  and Einstein's general relativity in entropic force approach. 
\end{abstract}
\vspace{2cm}
\section{Introduction}
The discovery of the profound connection between gravity and thermodynamics is one of the greatest achievements in theoretical physics \cite{2:1a,2:1b,2:2,2:3a,2:3b,newnew,2}. Hawking has derived that a schwarzschild black hole emits a thermal radiation whose temperature is given by $T=\frac{1}{8\pi M}$, where $M$ is the black hole mass \cite{2:3a,2:3b}. Bekenstein has shown that a black hole has a well defined entropy as $S=\frac{A}{4l_p^2}$, where $A$ is the area of the black hole horizon and $l_p$ is the Planck length. The black hole temperature, the horizon entropy and the black hole mass obey the first law of thermodynamics which can show that a black hole is a thermodynamical object \cite{1:3a,1:3b,1:3c}. In 1995, Jacobson showed that the Einstein field equations can be derived from the first law of thermodynamics together with the relation between the horizon area and the entropy. Following Jacobson, this novel idea has been investigated in modified theories of gravity \cite{1:6,1:13a,1:13b,1:14a,1:14b,1:5}. Recently, Verlinde proposed a new idea on the relation between gravity and thermodynamics \cite{1:16}. According to Verlinde's proposal, gravity is not a fundamental force and it can be considered as an entropic force arising from the change of information when a material body moves away from the holographic screen. Combining the entropic force with the Unruh temperature, the second law of gravitation can be obtained. In addition, combining the entropic force with the holographic principle and using the equipartition law of energy yield to Newton's law of gravitation. Recently, many works have been done to understand the entropic force \cite{2:12a,2:12b,2:12c,2:12d}.
In Verlinde's formalism the total number of bits is given by
\begin{eqnarray}
N=\frac{A}{G},
\end{eqnarray} 
which follows from the holographic principle and we have assumed $c=\hbar=1$. Since the Bekenstein-Hawking entropy-area relation is $S=\frac{A}{4l_p^2}$, we may have $N=4S$ by using $l_p^2=G$. It is clear that the entropy can be estimated as a measure of the microstates numbers in a system. Certainly, within the realm of high energy physics,  the candidates of the theory of quantum gravity can provide some corrections in the entropy-area relation and this yields to the modified $N$. 

Most of the promising candidates for quantum gravity expect the existence of a minimal observable length at the order of the Planck length \cite{minlength.1,minlength.2,minlength.3,minlength.4,minlength.5a,minlength.5b,minlength.5c}. Therefore, it is natural to take the planck length as a universal constant \cite{unilength}. On the other hand, length is obviously not an invariant under linear Lorentz boost. Therefore, the Lorentz symmetry at Planck scale may not be preserved. The goal of non-linear special relativity or doubly special relativity (DSR) is to preserve the relativity principle and at the same time treat Planck length as an invariant \cite{unilength3}. Within non-linear special relativity, the usual energy-momentum relation may be modified with some correction terms in the order of Planck length $l_p\simeq 1/{M_p}$ as
\begin{equation}\label{e-m r}
E^2f_1^2(l_pE)-p^2f_2^2(l_pE)=m_0^2,
\end{equation} 
where $f_1$ and $f_2$ are two general functions of energy with a constraint that they approach to unit for the energy scales much less than the Planck scale \cite{unilength4}. The modified dispersion relations may be responsible for threshold anomalies of ultra high energy cosmic rays and gamma ray burst \cite{1-10,1-11,1-12,1-13,1-14,1-15,1-16} and contribute corrections to the black hole thermodynamics \cite{unilength}.
Recently, non-linear special relativity has been generalized to incorporate the effects of gravity, leading to the rainbow gravity model. In rainbow gravity, the metric of the background detected by any probe is not fixed, but depends on the energy of the probe \cite{unilength5}. The rainbow gravity can be used to study the black holes. Of course, the thermodynamical properties of the black holes may be influenced by the effects of rainbow gravity. One can use the corrected entropy-area relation within rainbow gravity to find the modified $N$. 

In this paper, we will study the black hole thermodynamics within rainbow gravity. The corrected entropy-area relation leads to the modified $N$. Using the modified $N$, we will study the effects of rainbow gravity in the theory of modified Newtonian dynamics (MOND). We will also obtain the corrections to the gravitational force and gravitational potential. Considering the relativistic effects, the Einstein's field equations will also be modified.
 
\section{The black hole thermodynamics within rainbow gravity }
The metric of a Schwarzschild black hole in rainbow gravity model can be written as
\begin{equation}\label{metric}
ds^2=-\frac{(1-\frac{2GM}{r})}{f_1^2}dt^2+\frac{(1-\frac{2GM}{r})^{-1}}{f_2^2}dr^2+\frac{r^2}{f_2^2}d\Omega^2,
\end{equation}
which is the spherically symmetric solution of $G_{\mu \nu}(E)=8 \pi G(E) T_{\mu\nu}(E)$ and $G$ is the Newton's constant. 
The position of the horizon is $r_+=2GM$.
Since we are interested in studying gravity as an entropic force and modifying it within rainbow gravity, we are going to derive the corrections to black hole thermodynamics from rainbow gravity.
The surface gravity $\kappa$ on the black hole horizon can be obtained by 
\begin{equation}\label{T-kappa}
\kappa=-\frac{1}{2} \lim_{r  \rightarrow r_+} \sqrt{-\frac{g^{11}}{g^{00}}}\frac{(g^{00})'}{g^{00}}.
\end{equation} 
Then the surface gravity of a Schwarzschild black hole in rainbow gravity is
\begin{equation}\label{T-kappa+}
\kappa=\frac{f_2}{f_1}\frac{1}{4GM},
\end{equation}
and it is possible to find the temperature from the surface gravity by $T=\frac{\kappa}{2\pi}$ \cite{rainbow}.
As a result, the modified temperature with the effects of rainbow gravity can be written as
\begin{equation}\label{temperature}
T=\frac{f_2}{f_1} \frac{1}{8\pi GM}. 
\end{equation} 
It can be seen that the temperature of the black hole is different for probes with different energies.
One can write the temperature as 
\begin{equation}\label{TT0}
T=\frac{f_2}{f_1} T_0, 
\end{equation}
where
\begin{equation}\label{T}
T_{0}=\frac{1}{8\pi GM}.
\end{equation}
$T_0$ is the black hole entropy without any corrections from rainbow gravity and $T$ is the temperature which is modified by taking into account the effects of $f_1$ and $f_2$. 

By identifying probes with radiation particles in the vicinity of the horizon of the black hole, we are able to define an intrinsic temperature for large modified black holes. Assuming that the dominant part of the radiation particles are photons with average energy $E=<E>$ and $m_0=0$, the temperature of the black hole can be identified with the energy of the photons emitting from the black hole as $T \simeq E$ \cite{rainbow}.

It is necessary to introduce the specific forms of the functions $f_1$ and $f_2$ to find the black hole temperature. According to \cite{unilength,rainbow,sefiedgar2}, we can write
\begin{equation}\label{f_1 and f_2}
(f_1)^2=1-\alpha l_p^2E^2, \qquad (f_2)^2=1,
\end{equation}
where $\alpha$ is a positive quantity of order one which have been input to distinguish the correction terms arising from the rainbow gravity effects.  For $\alpha=0$, the energy-momentum relation reduces to its standard form in special relativity.  
By the definitions of $f_1$ and $f_2$ and plugging $T\simeq E$, equation (\ref{TT0}) yields to 
\begin{equation}
{T}^2=\frac{1}{1-\alpha l_p^2{T}^2}T_0^2,
\end{equation}
which can be solved to find the Schwarzschild black hole temperature as
\begin{equation}\label{TTTT}
{T}=\left[\frac{1-\sqrt{1-4\alpha l_p^2T_0^2}}{2\alpha l_p^2} \right]^{\frac{1}{2}}.
\end{equation}
For large black holes with $4\alpha l_p^2T_{0}^2 \ll 1$, the modified temperature reduces to the temperature of an ordinary Schwarzschild black hole, $T \simeq T_{0}$. For small black holes with extremely high temperature, the temperature reaches its maximal value $T_{max} \simeq \frac{1}{\sqrt{2\alpha}l_p}$ as $T_{0} \simeq \frac{1}{2\sqrt{\alpha}l_p}$. Correspondingly the radius of the black hole horizon is bounded from below by $r_+ \geq \frac{\sqrt{\alpha}l_p}{2\pi}$. The existence of a minimum radius leads to the possibility of the existence of the black hole remnant at the end of the evaporating process which can be considered as a suitable candidate for dark matter.
Using $A=4\pi r_+^2=16\pi G^2M^2$, the first law of thermodynamics, $dM=TdS$, can be applied to find the entropy-area relation within rainbow gravity as
\begin{eqnarray}\label{entropy}
S=\frac{A}{4l_p^2}-\frac{\alpha}{32\pi} \ln {\frac{A}{4l_p^2}}+\frac{5\alpha^2 l_p^2}{512 \pi^2}\frac{1}{A}.
\end{eqnarray}  
Without any loss of generality in conclusions, we have only considered the correction terms up to the second power of $\alpha$ in equation (\ref{entropy}). The appearance of the logarithmic correction term in the entropy-area relation is consistent with the results obtained in string theory and loop quantum gravity \cite{unilength-26,unilength-27,unilength-28,unilength-29}. 

\section{The corrections to the modified Newtonian dynamics (MOND)}
As an alternative to non baryonic dark matter, Milgrom proposed the new idea of modified Newtonian dynamic in 1983 \cite{newnew,2,2:44a,2:44b,2:44c}. To solve the problem of mass discrepancies in the galaxy rotation curves, he founded the necessity of the modifications to Newtonian dynamics  for acceleration smaller than $1.2 \times 10^{-10}m/s^2$. Considering the Newtonian acceleration, $a_N$, and $a_0=1.2\times10^{-10}m/s^2$, one can find the acceleration due to gravity as $a=\sqrt{a_Na_0}$. The MOND acceleration due to gravity can be found from the relation
\begin{eqnarray}
a_N=a\mu\left(\frac{a}{a_0}\right).
\end{eqnarray} 
The interpolation function $\mu\left(\frac{a}{a_0}\right)$ admits $\mu=1$ for $a \gg a_0$ and  $\mu=\frac{a}{a_0}$ for $a \ll a_0$ \cite{newnew,2,2:45}.
MOND can  be derived from the holographic entropy area relation \cite{2:40} and the collective motion of holographic screen bits \cite{2:48}, where bits are units of information on the holographic screen. In the critical phenomena of cooling, it can be shown that the notion of MOND can be derived if the zero energy bits are removed from the total number in the equipartition relation. But a modified equipartition theorem should be applied which assumes that the division of energy is not homogeneous on all bits below a critical temperature. Then, the theory of MOND can be recovered along with the holographic principle and the Unruh temperature \cite{newnew,2,2:49}. The fraction of bits with zero energy is given by
\begin{eqnarray}
N_0=N\left(1-\frac{T}{T_c}\right).
\end{eqnarray}
There are no bits with zero energy for $T \geq T_c$ and the zero energy phenomena starts for $T<T_c$. The number of bits with non zero energy is given by
\begin{eqnarray}
N-N_0=N\left(\frac{T}{T_c}\right).
\end{eqnarray}
Using the equipartition law of energy, we can write
\begin{eqnarray}
E=\frac{1}{2}\left(N\frac{T}{T_c}\right)T,
\end{eqnarray}
where we have put $k_B=1$. By defining $M$ as an emergent mass which is at the center of the space enclosed by the holographic screen and using $E=M$, we can write
\begin{eqnarray}\label{T2holography}
T^2=\frac{2MT_c}{N}.
\end{eqnarray}
By utilizing the Unruh temperature, $T=\frac{a}{2\pi}$, which is the temperature experienced by an observer in an accelerated frame, equation (\ref{T2holography}) can be written as
\begin{eqnarray}\label{Na2}
Na^2=8\pi^2 M T_c.
\end{eqnarray}
It is now useful to obtain the modified $N$ from the corrected entropy in equation (\ref{entropy}) as 
\begin{eqnarray}\label{N=4S}
N=\frac{A}{l_p^2}-\frac{\alpha}{8\pi} \ln {\frac{A}{4l_p^2}}+\frac{5\alpha^2 l_p^2}{128 \pi^2}\frac{1}{A},
\end{eqnarray}
where we have utilized the relation $N=4S$.
Using  equation (\ref{N=4S}) and the relation $A=4\pi R^2$, one can rewrite equation (\ref{Na2}) as 
\begin{eqnarray}
a^2\frac{4\pi R^2}{l_p^2}\left[1-\frac{\alpha l_p^2}{32\pi^2 R^2} \ln {\frac{\pi R^2}{l_p^2}}+\frac{5\alpha^2 l_p^4}{2048 \pi^4 R^4} \right]=8\pi^2 MT_c.
\end{eqnarray}
By considering $a_0=2\pi T_c$, one can write
\begin{eqnarray}
a\left(\frac{a}{a_0}\right)=\frac{Ml_p^2}{R^2}\left[1+\frac{\alpha l_p^2}{32\pi^2 R^2} \ln {\frac{\pi R^2}{l_p^2}}-\frac{5\alpha^2 l_p^4}{2048 \pi^4 R^4} \right],
\end{eqnarray}
which is the modified Newtonian dynamics obtained within rainbow gravity.

\section{The corrections to the Newton's law of gravitation}
To derive the Bekenstein's entropy-area relation, it has been assumed that a particle is a part of the black hole if it is one Compton wavelength from the horizon. Then, there will be a change in mass and area of a black hole which is defined as a bit of information. Motivated by Bekenstein's argument, Verlinde postulated that the change of entropy associated with the information on the boundary equals
\begin{eqnarray}
\Delta S=2\pi,  \qquad  \Delta x=\frac{1}{m},
\end{eqnarray}
where $k_B=c=\hbar=1$ \cite{1:16}.  
Assuming a linear relation between the entropy changes and the displacement $\Delta x$, one can write
\begin{eqnarray}\label{I}
\Delta S=2\pi m \Delta x.
\end{eqnarray}
To reveal the entropic force, Verlinde used the analogy with osmosis across a semi-permeable membrane. When a particle has an entropic reason to be on one side of the membrane and membrane carries a temperature, it will experience an effective force equal to
\begin{eqnarray}\label{II}
F \Delta x=T \Delta S.
\end{eqnarray}
To have a non zero force, it is necessary to have a non vanishing temperature. Furthermore, one can find from Newton's law that a force leads to a non zero acceleration. Therefore, one can conclude that acceleration and temperature are closely related. The relation between acceleration and temperature can be obtained via Unruh effect.
Assuming that the total energy $E$ of a system is divided evenly over $N$ bits, then the equipartition law of energy can be written as
\begin{eqnarray}
T=\frac{2E}{N}.
\end{eqnarray} 
Using $E=M$, one can write
\begin{eqnarray}\label{III}
T=\frac{2M}{N}.
\end{eqnarray}
From equations (\ref{I}), (\ref{II}) and (\ref{III}) one can obtain
\begin{eqnarray}\label{F}
F=\frac{4\pi m M}{N}.
\end{eqnarray}
Substituting $N$ from equation (\ref{N=4S}), the force can be written as
\begin{eqnarray}\label{F(R)}
F=\frac{GmM}{R^2}\left[ 1+\frac{\alpha G}{32\pi^2 R^2} \ln {\frac{\pi R^2}{G}}-\frac{5\alpha^2 G^2}{2048 \pi^4 R^4}    \right],
\end{eqnarray} 
where $A=4\pi R^2$ has been used and $G=l_p^2$. This is the modified version of the gravitational force which is obtained from rainbow gravity. When $\alpha=0$, the modified force is reduced to the usual Newton's law.
In addition, the Newtonian potential can be obtained from the force as
\begin{eqnarray}\label{V(R)}
V=\frac{GmM}{R}\left[ 1+\frac{\alpha G}{96\pi^2 R^2} \ln {\frac{\pi R^2}{G}}+\frac{\alpha G}{144 \pi^2 R^2}-\frac{\alpha G^2}{2048\pi^4 R^4}    \right].
\end{eqnarray} 
One can find that the gravitational potential admits some corrections from rainbow gravity. It is clear that the effects of correction terms are important in short scales and not in large scales. In large scales, the contribution of the Newtonian term is dominant. The necessity of modifying the gravitational potential at short scales has also been verified in the previous works \cite{newnew,2,2:12d,callin,Bronnikov,Jung,tawfik,awad}.  The absence of the correction term proportional to $\frac{1}{R^2}$ in the gravitational potential is originated from choosing $f_1^2=1-\alpha l_p^2E^2$ in the modified dispersion relation. If one insists on the emergence of a correction term proportional to $\frac{1}{R^2}$, the choice of $f_1^2=1+\beta l_pE-\alpha l_p^2E^2$ and $f_2^2=1$ can be helpful.

\section{The corrections to Einstein's field equations}
It is possible to extend the derivation of the laws of gravity to the relativistic case and obtain Einstein's field equations \cite{1:16}. Now we can consider a holographic screen which is enclosing a certain static mass configuration with total mass $M$. The bit density can be derived from equation (\ref{N=4S}) as 
\begin{equation}\label{dN}
dN=\frac{dA}{G}\left[ 1-\frac{\alpha G}{8\pi A}-\frac{5\alpha^2 G^2}{128\pi^2 A^2}\right].  
\end{equation}
Let us assume that the energy associated with $M$ is distributed over all the bits and each bit carries a mass unit equal to $\frac{1}{2}T$ due to the equipartition law. Hence
\begin{equation}\label{M1}
M=\frac{1}{2} \int_{\mathcal{S}} TdN,  
\end{equation}
and the local temperature on the screen is given by
\begin{equation}\label{T1}
T=\frac{1}{2\pi} e^{\phi} n^b \nabla_b \phi.  
\end{equation}
It is necessary to point that $e^{\phi}$ is the redshift factor as $T$ is measured by an observer at infinity \cite{2}. Substituting equations(\ref{dN}) and (\ref{T1}) into (\ref{M1}), on can obtain
\begin{equation}\label{M}
M=\frac{1}{4\pi G} \int_{\mathcal{S}} e^{\phi} \nabla \phi. \left[ 1-\frac{\alpha G}{8\pi A}-\frac{5\alpha ^2 G^2}{128\pi^2 A^2} \right]dA.  
\end{equation}
Equation (\ref{M}) is the modified Gauss law in general relativity. The right hand side of the equation is the modified Komar mass while the first integral on the right is precisely komar 's definition of mass which is contained inside a volume in a static curved space time
\begin{equation}\label{MK}
M_K=\frac{1}{4\pi G} \int_{\mathcal{S}} e^{\phi} \nabla \phi . dA .
\end{equation}  
According to \cite{2,1:16}, Komar mass can be expressed in terms of Ricci tensor $R_{ab}$ and the Killing vector $\xi^{a}$ by using the Stokes theorem and the killing equation, $\nabla^a \nabla_a \xi^b=-R^b_a\xi^a$. Hence
\begin{equation}\label{MK+}
M_K=\frac{1}{4\pi G}\int_{\Sigma}R_{ab}n^a\xi^b dV.
\end{equation}
On the other hand, one can write $M$ as a volume integral of the stress energy tensor $T_{ab}$ as
\begin{equation}\label{MMMM}
M=2\int_{\Sigma} \left(T_{ab}-\frac{1}{2}Tg_{ab}\right)n^a\xi^b dV.
\end{equation}
As a result, one can use the equations (\ref{MK+}) and (\ref{MMMM}) to rewrite equation (\ref{M}) as
\begin{equation} 
\int_{\Sigma} \left[R_{ab}-8\pi G (T_{ab}-\frac{1}{2}Tg_{ab})\right]n^a\xi^b dV=-\int_{\mathcal{S}} e^{\phi}\nabla \phi .\left(-\frac{\alpha G}{8\pi A}-\frac{5\alpha^2 G^2}{128\pi^2 A^2}\right)dA,
\end{equation}
where $\Sigma $ is the volume bounded by the holographic screen $\mathcal{S}$ and $n^a$ is the normal.
Using $F=-me^{\phi}\nabla \phi$ in \cite{1:16}, one can find by some manipulations that
\begin{equation}
R_{ab}=8\pi G\left(T_{ab}-\frac{1}{2}Tg_{ab}\right)\left[1-\frac{\alpha G}{8\pi A}-\frac{5\alpha^2 G^2}{256\pi^2 A^2}\right],
\end{equation}
which is the modified version of Einstein field equations. It is clear that the field equations and the geometry of the space time are affected by rainbow gravity and they are observer dependent. When $\alpha=0$, the field equations reduce to the Einstein field equations. 

As an important point, one can find that all the results obtained from rainbow gravity are approximately comparable with ones derived via generalized uncertainty principle \cite{2,tawfik,awad}. Of course one can find the same functional results from both approaches by choosing the model parameters in a suitable manner. What is the importance of the comparison of two approaches? The dependence of the results on the energy of the probe in rainbow gravity and the existence of the minimal observable length as an origin of the generalized uncertainty principle are obviously related to each other in this comparison. 

\section{Conclusions}
Rainbow gravity as a model in which the metric is energy dependent has been considered to study the thermodynamical properties of a  Schwarzschild black hole. One can find that the black hole thermodynamics admits some corrections from the effects of rainbow gravity. Since the black hole entropy can be used to find the number of bits $N$, the corrections to the entropy of a black hole are necessary to be obtained. The importance of deriving a good estimation of $N$, as a measure of the number of microstates, is clear in the approach of Verlinde who assumes gravity as an entropic force. In this paper, the entropy in rainbow gravity has been corrected and it has been used to derive the modified $N$. The modified $N$ has lead to the corrections to the modified Newtonian dynamics. The effects of rainbow gravity on the gravitational force and gravitational potential have also been obtained. It has been found that it is necessary to take into account the corrections to gravitational force and gravitational potential in short scales, since the contribution of the correction terms are more important there. By extending the approach to the relativistic case, the Einstein's field equations have also been modified. The dependence of the modified field equations on the energy of the probe according to the appearance of $\alpha$  may be interesting to be studied more.

\end{document}